\documentclass{elsart}
\usepackage{epsfig}
\newcommand{\avg}[1]{\langle {#1} \rangle}
\begin{document}
\begin{frontmatter}
\title{Filtering information in a connected network}
\author[frib]{Andrea Capocci},
\author[prague]{Franti\v{s}ek Slanina}, and
\author[frib]{Yi-Cheng Zhang}

\address[frib]{Institut de Physique Th{\'e}orique,
Universit{\'e} de Fribourg, Perolles CH-1700, Switzerland}
\address[prague]{
        Institute of Physics,
    Academy of Sciences of the Czech Republic,\\
    Na~Slovance~2, CZ-18221~Praha,
    Czech Republic\\
        e-mail: slanina@fzu.cz
}
\begin{abstract}
We introduce a new kind of Information Theory. From a finite
number of local, noisy comparisons, we want to design a robust
filter such that the outcome is a high ranking number, Both
analytical and numerical results are encouraging and we believe
our toy model has wide ranging implications in the future 
Internet-based information selection mechanism.
\end{abstract}


\end{frontmatter}
\section*{Introduction}

Claude Shannon, more than half a century ago, has introduced the
modern Information Theory \cite{Sh}. This theory was a timely
contribution to the the explosive growth of long distance
communications during, or just after, war time. He studied, in a
general framework, the coding strategies that allow one to send a
message through a noisy channel, such that the received signal
contains no error. Actually, it is an easy task, if one is allowed
unlimited {\em information capacity}, that is, infinite number of
bits transmitted through the channel in a time unit. For instance,
one could repeat the same binary signal many times. On the
receiving end, taking the average or using the majority rule
\cite{McK}, the original message can be restored to any desired
precision.

But, unlike in academic economics literature, unlimited information
resources are not available. Therefore, one usually deals with the
problem of minimizing the information capacity needed to transmit
messages with a required degree of precision. Indeed, one can code
a message in an efficient way such that only a very limited amount
of redundancy is needed. For this purpose, Shannon proved two
fundamental theorems and found a lower bound on the necessary
redundancy. However, his theory does not provide a constructive
method to code signals. In other words, no methods
were introduced by Shannon to solve the problem, except in some very
special cases. Since then, a number of coding methods have been
built, that reach the optimal efficiency computed by Shannon.
Indeed, much of the state-of-art of the information theory is
dedicated to sophisticated coding/decoding business \cite{McK}.

The current Internet era poses new challenges. Nowadays, the
problem does not lie in the lack of information, but in the fact
that too much of it is available. One therefore has to set up
methods to sort the specific information out of a highly
disordered environment according to any given relevance criteria. Having
infinite resources, in principle one could examine in depth each
available information source, selecting the relevant ones.
Unfortunately, one does not have such resources. Many recent 
 researches emphasize that scarcity does not affect 
available information any longer, but rather the 
capability to process it. In other words, the most valuable good of
the new economic era is {\em attention} \cite{Att}.
Thus, in a finite time one can obtain at most an approximate 
estimate of the true relevance of all available information 
sources.

In addition, in many common cases the quality of an information
source cannot be observed directly, but only through comparison
with peers. For example, one cannot measure the
intrinsic relevance of a web page by simply observing it: most of
the times, one needs a collection of other pages to compare
their content, in order to choose the best.
Unfortunately, the results of such comparisons are often fuzzy,
since a clear-cut assessment would need prohibitive amount of
time and attention resources. Nevertheless, a certain number of
matches allows one to have an approximate estimate of the
intrinsic relevance. Clearly, as the number of matches is increased,
the accuracy in the observation grows.

Finally, the results of a search have to be presented in a shape
which takes into account the human interaction. For example, new
generation WWW search engines put a strong effort in establishing
reliable {\em rankings} of relevant web pages matching any given
query \cite{Google}. 
Instead of showing a whole rankings, some search engine provide a 
restriced number of matches to a query, or even a single one, 
as in Google's ``I'm feeling lucky'' option.
Again, this option implies a neat gain in ease at use, though it rarely 
corresponds to the best possible answer. 

In this paper we propose a toy model of information filtering. Our
model deals with the problem of finding the most relevant element
from a large set, in presence of a stochastic {\em noise} which
prevents a perfect perception of the intrinsic quality of
each item. For the sake of simplicity, we represent the quality of
each item by a real number randomly drawn from a uniform distribution in
the range $[0,1]$.

Knowing all the qualities, it would be trivial to sort the items
and build a ranking: in this case, the higher the quality of an
item, the higher its rank. But, as explained above, we assume that
these numbers cannot be accessed directly: any information can be
obtained by a noisy comparisons among the items. In particular,
we choose a pairwise filtering architecture, i.e. information can
be gathered only by comparing pairs of items.

For a real life instance, one could think to football teams in a
national league. To choose the best team, we cannot restrict
ourselves to examine each team and draw a judgment: many people do
so, without ever reaching an ultimate answer. On the other hand,
we can guess an approximate answers, using a finite number of
pairwise matches. Actually, the outcome of a match does not signal
precisely a better intrinsic quality of one team over the other. But if a
team is intrinsically better than another one, it has a greater
chance of winning in a real game. If an infinite number of
games was allowed, we would be able to discern the intrinsic
superiority by pitting two teams against each other.

In reality, we must be content with approximate answers, using
rather a limited number of comparisons (in the soccer language,
number of games). Clearly, this introduces a design problem. The
aim is to get as good an approximation using a given number of
comparisons; or, equivalently, to achieve a given level of
approximation using minimalizing necessary resources. In this
sense, we say our approach is a generalization of Shannon's
Information Theory.

From sport events, we learn that a different structure gives rise
to a different quality of filtering. In the final round of the Soccer
World Cup, for example, teams are disqualified after a single
defeat, up to the final match. This architecture does not often
yield a very good approximation of intrinsic quality ranking,
because a good team defeated by a worse team will be put out of
the competition. But this tree structure is the msot economical
one, needing a minimal number of games.

On the other hand, in european national leagues each team plays
each other. This structure yields in general a more reliable
approximation of the intrinsic ranking, since more redundancy is built into
the scheme, as teams are not menaced from chancy elimination. 
But, as in
Shannon Information theory, more precision requires more
resources, and a trade-off has to be made between the two
considerations. In our metaphor, the World Cup need less time
resources, usually one month, whereas many national leagues take
almost a year. This example illustrates the dilemma.

We study a very simple design structure which allows both
elimination of low quality items, which decreases the time needed
to the final selection, and a certain degree of redundancy, which
provides reliability to our mechanism.

\section*{The model}

The underlying structure of our filtering model is a
one-dimensional lattice of $L$ nodes with periodic boundary
conditions. Time is assumed to be an integer variable
$t=0,1,2,\ldots$. Every node $i=1, \ldots,L$ is attached a value
$x_{i}(t)$. The starting configuration is an array
${x_i(t=0)}_{i=1,\ldots,L}$ of random variables drawn from the
range $[0,1]$ with uniform probability: these values represent the
intrinsic qualities we introduced above. At each time step, every
couple of neighbors values $(x_i(t),x_{i+1}(t+1))$ for
$i=1,\ldots,L$ gives rise to a new value $x_i(t+1)$, according to
the following rule:
\begin{displaymath}
x_{i}(t+1)= \left\{
\begin{array}{ll}
x_{i}(t) & \textrm{with probability}
\frac{x_{i}(t)}{x_{i}(t)_+x_{i+1}(t)}
\\
x_{i+1}(t) & \textrm{with probability}
\frac{x_{i+1}(t)}{x_{i}(t)_+x_{i+1}(t)}
\end{array}
\right \}
\end{displaymath}
Therefore, in a time step the whole array is updated. The periodic
boundary conditions ensure that $x_{L+1}=x_{1}$ for all values of
$t$, and thus consistency of the evolution law.

\begin{figure} \label{fig0}

\centerline{\epsfxsize=5.in \epsffile{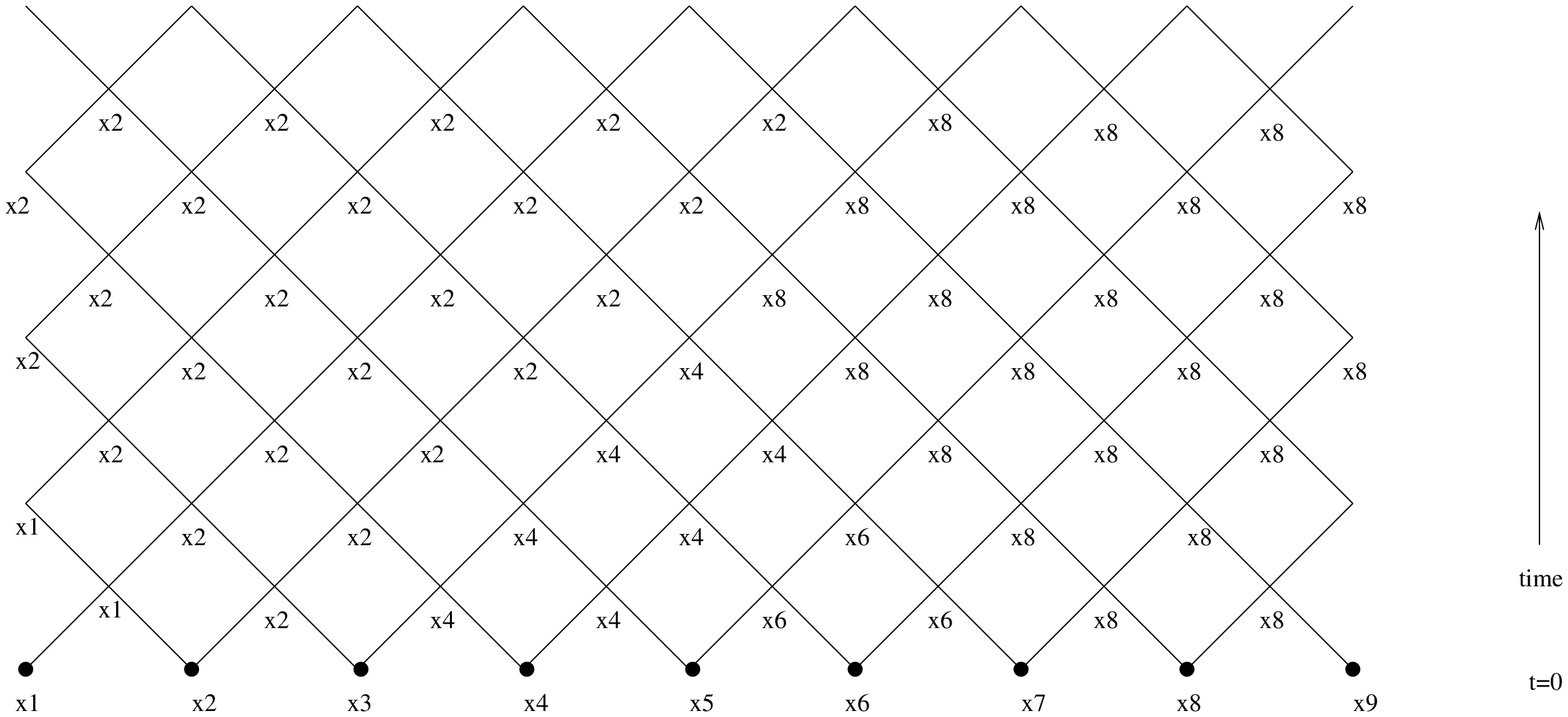}}
\caption{
A realization of the model.
The vertical direction represents
the time evolution.
}
\end{figure}

As the chain evolves, connected {\em domains}, i.e. regions of the
lattice composed of sites occupied by the same value, emerge: high
values have a greater probability to spread over neighbor sites,
and low values are more likely to vanish, as described in figure
\ref{fig0}.
Eventually, after a time $t^{*}$, all sites are occupied by a
single domain associated to the value $x^*$: the process has
reached a stationary state. Accordingly, we call this system a
{\em filter}, since it exerts a selection on the initial $L$
values, favoring by its dynamics the propagation of domains
associated with high values.

The value $x^*$ is attached to a given site $s$ in the starting
configuration, i.e. $x^* \in {x_i(0)}_{i=1,\ldots,L}$. We say that
the site $s^{*}$ and the value $x^{*}=x_{s^*}(0)$ associated with
it at the beginning have been {\em selected} by the filter. We
define the {\em search} time $t^{*}(L)$ needed to reach the
stationary state, and the $inefficiency$ of this filter, i.e. the
rank $R(L)$ of the selected value in the starting configuration.
An ideal filter would select the site associated to the highest
value in the starting configuration ($R(L)=1$). Due to the
randomness in the initial condition and the stochastic dynamics,
our filter may select a different site. We will investigate these
two quantities, $t^*(L)$ and $R(L)$, as a function of the total
number of sites $L$.

The number of {\em different} values in the lattice monotonically
decreases with time. We denote by $n(t,L)$ the number of these
domains. A relation between $t$, $L$ and $n$ can be derived by
simple reasoning. Let us assume that at time $t(n,L)$ only $n$
domains remain in a chain of length $L$.

Each of them occupies, on average, a region of size $\frac{L}{n}$.
Therefore, $t(n,L)$ approximately corresponds to the time needed
to reach the stationary state for a filter acting over this
sub-region of size $\frac{L}{n}$. This reads
\begin{equation} \label{hier1}
t(n,L)=t(1,\frac{L}{n}).
\end{equation}
By assuming
$n(t,L) \sim t^{-\alpha}L^{\beta}$,
with $\alpha > 0$ and $\beta >0$, we write
$t(n,L) \sim n^{-1/\alpha}L^{\beta/\alpha}$, and replace this
expression in eq. \ref{hier1},
This way, we obtain
\begin{equation}
n^{-\beta/\alpha}L^{\beta/\alpha} = n^{-1/\alpha}L^{\beta/\alpha},
\end{equation}
which implies $\beta = 1$ and $t^{*}(L) \sim
L^{\frac{1}{\alpha}}$. This rough estimate of the scaling behavior
with respect to $L$ is confirmed by the analysis of the population
dynamics of the coalescing domain walls.

\section*{Search time}
\begin{figure} \label{fig1}

\centerline{\epsfxsize=5.in \epsffile{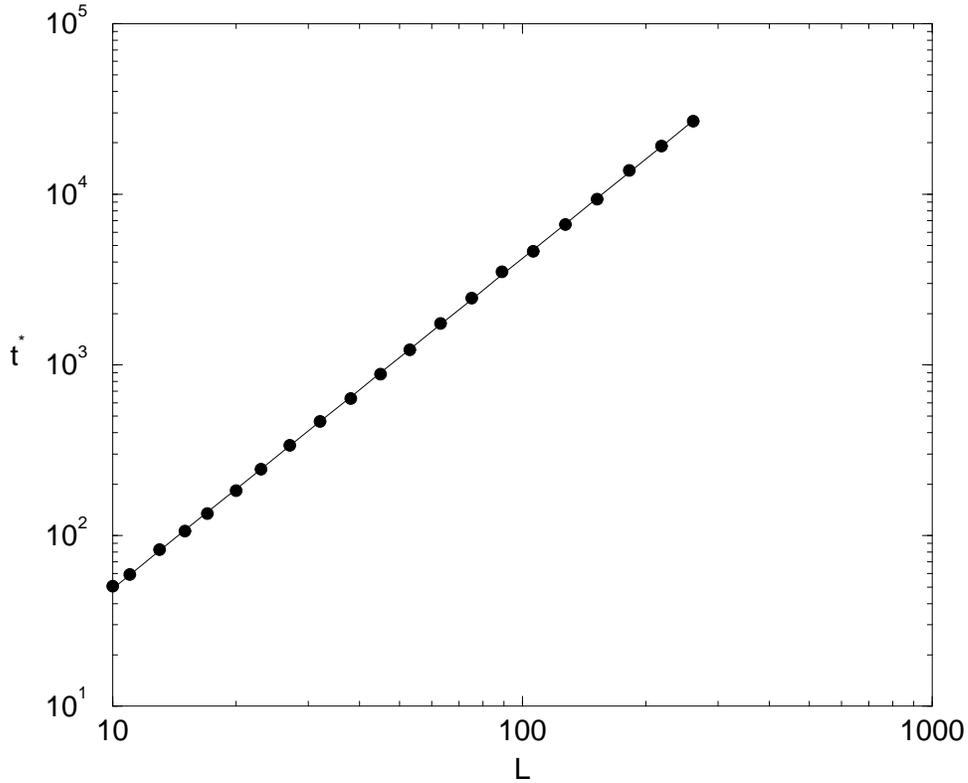}}
\caption{
Search time as a function of the total number of sites $L$.
Circles correspond to numerical simulations.
The solid line has slope $1.94$ in log-log scale.
}
\end{figure}

At time $0$, there are $L$ domain walls, since each domain is made
of a single site. As time passes by, domains vanishes. A domain
vanishes when the two surrounding domain walls coalesce. By
tracing all the domain walls as a function of time, following the
framework of $1+1$ dimensional directed polymers, we observe a
tree-like structure, whose source is in the end point of the
filter process. At time $t$ the lattice is occupied by $n(t,L)$
domains $\Gamma_{k}$, corresponding to the values
${y_{k}=x_{i_{k}}(0)}$, with $k=1,\ldots,n(t,L)$. Let us denote by
$h_{k}(t)$ the position of the $k$-th wall between domains
$\Gamma_{k}$ and $\Gamma_{k+1}$. The wall between these domains
performs a random walk of unitary steps whose {\em drift} $v(n)$
is equal to $\frac{y_{k+1}-y_{k}}{y_{k+1}+y_{k}}$. To evaluate the
time $\Delta(n)$ between two subsequent coalescing along the same
domain wall, when the surviving domain walls are $n$, we write
\begin{equation} \label{delta}
\Delta(n) \sim \frac{L}{n\avg{|v(n)|}},
\end{equation}
where $L/n$ is the average length of a domain, and $\avg{|v(n)|}$
is the typical speed with which a wall moves towards its neighbor,
and the average is performed over the distribution of the
remaining values $\{y_{k}\}_{k=1,\ldots,n}$ at time $t$. Therefore
$\Delta(n)^{-1}$ is the probability per time step that a domain
wall encounters a neighbor. Since there are $n$ walls, the total
probability of a coalescing anywhere in the lattice is
$n/\Delta(n)$. At each intersection, the number of walls decreases
by one, therefore we can write a differential equation for
$n(t,L)$,
\begin{equation} \label{n_t}
\dot{n} = -\frac{n}{\Delta(n)}.
\end{equation}
We can give an estimate of $\avg{|v(n)|}$ by assuming that the
values ${y_{k}}_{k=1,\ldots,n}$ are uniformly distributed in the
range $[1-n/L,1]$. Under this assumption,
\begin{eqnarray}
\avg{|v(n)|} & = & \frac{L^2}{n^2}\int_{1-\frac{n}{L}}^{1}dx \int_{1-\frac{n}{L}}^{1}dy \frac{|x-y|}{x+y} = \\
         & = & \frac{1}{2}[-\frac{2n}{L}+\frac{n^2}{L^2}+2 \ln{2} + 2(\frac{n^2}{L^2} - 2)\ln{2-\frac{n^2}{L^2}} -  \\
             & - & 2(\frac{n}{L}-1)^2\ln{2(1-\frac{n}{L})}].
\end{eqnarray}
If the remaining domains are $n \ll L$, we can expand this
expression in a series of powers up to the first order in
$\frac{n}{L}$, getting
\begin{equation} \label{drift}
\avg{|v(n)|} \simeq \frac{n}{12L} + \ldots.
\end{equation}
By replacing eq. (\ref{drift}) in eq. (\ref{delta}), we obtain
$\Delta \sim \frac{L^{2}}{n^{2}}$.
Then, for $n \ll L$, the time evolution of $n(t,L)$ reads
$\dot{n} \sim n^{3}L^{-2}$.
Given the assumption made on the scaling behavior
of $n$ with respect to $t$ and $L$, the previous equation
provides us with two relations for $\alpha$ and $\beta$,
which yield $\alpha=\frac{1}{2}$ and $\beta = 1$.
By replacing the steady state condition
$n(t^{*}) = 1$ in the scaling relation of $n$ as a
function of $t$ and $L$, we obtain
$(t^{*})^{-\frac{3}{2}}L \sim L^{-2}$.
This gives the scaling relation
\begin{equation} \label{t*_L}
t^{*} \sim L^{2},
\end{equation}
which is confirmed by numerical simulation, as shown
in fig. \ref{fig1}.
\section*{Inefficiency}
To compute the inefficiency of the filter, we have to estimate the
rank of the finally selected value in the starting configuration.
Let us make some strong (but reasonable) approximation: (i) we
assume that the distribution $\psi(y)$ of the remaining $n(t)$
numbers on the lattice is uniform for all times, with mean value
$\bar{y}$ and support $[2\bar{y}-1, 1]$. (ii) we make a mean field
approximation about the time evolution of the domains. In order to
explain better hypothesis (ii), let us focus on a single domain.
Let $y_k$ be the value occupying a domain, $l_k(t)$ the length of
this domain, $y_{k-1}$ and $y_{k+1}$ the values occupying the
neighboring domains. The time evolution of $l_k$ reads
\begin{equation} \label{timev}
l_k(t+1) = l_k(t) - \frac{1}{2}[\eta(t,y_{k-1},y_{k}) -
\eta(t,y_k,y_{k+1})]
\end{equation}
where
\begin{displaymath}
\eta(t,x,y) = \left\{
\begin{array}{ll}
1 & \textrm{with probability} \frac{x}{x+y}
\\
-1 & \textrm{with probability} \frac{y}{x+y}
\end{array}
\right
\}
\end{displaymath}

We now replace in eq. (\ref{timev}) the neighboring values $y_{k
\pm 1}$ by the mean value of the distribution of remaining
numbers, $\bar{y}$. This assumption is equivalent to considering
each domain size as a biased random walk starting at position
$l_k(0)$ with an absorbing boundary at the origin (that
corresponds to the coalescence of two neighboring domain walls).
The bias of the random walk is given by the interaction between
the domain occupied by the value $y_k$ and the effective medium,
which is assumed to be occupied by the value $\bar{y}$. In the
same mean field approach, we assume that $l_k(0)=L/n(t)$, which is
the typical distance between two adjacent walls. Such a random
walk \cite{Fel} is absorbed by the boundary with probability $1$
if $y_k < \bar{y}$ and probability $q(y_{k}) = \biggl(
\frac{\bar{y}}{y_k} \biggr) ^{2L/n(t)}$ otherwise. In both cases,
the absorption occurs after the same typical time $t_0$.

\begin{figure} \label{fig2}

\centerline{\epsfxsize=5.in \epsffile{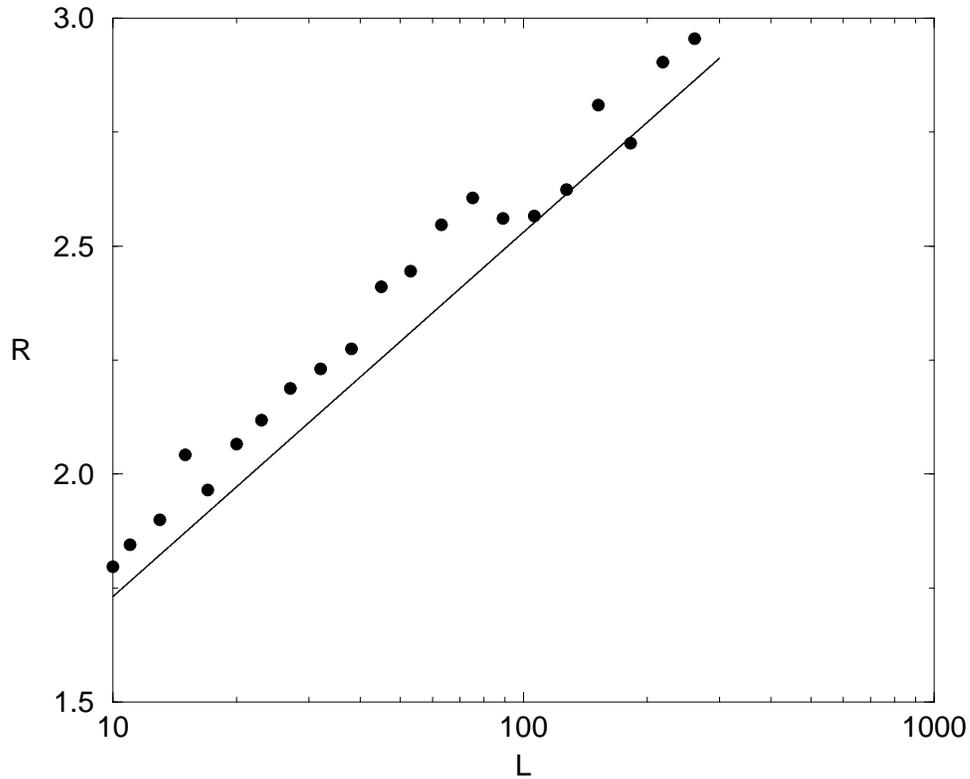}}
\caption{
Inefficiency of the filter as a function of the total number of sites $L$.
Circles correspond to numerical simulation; the solid line corresponds to
eq. \ref{rank}
}
\end{figure}

After $m$ typical times (not all equal!), the mean value of the
distribution is $\bar{y}(m) = 1-2^{-m-1}$. In fact, in our
approximation, during a typical time all domains occupied by
values lower than $\bar{y}$ have vanished, as does a fraction of
the domains occupied by values higher than $\bar{y}$. This
fraction is estimated by
\begin{eqnarray}
q(m) & = & \int_{\bar{x}}^{1} \psi(y) \biggl( \frac{\bar{y}}{y} \biggr) ^{2L/n(m)} dy = \\
     & = & \frac{\bar{y}^{2L/n(m)}}{1-\bar{y}} \int_{\bar{x}}^{1} \frac{1}{y^{2L/n(m)}} dy = \\
     & = & (1-2^{-m-1})^{2L/n(m)} + \ldots = \\
     & \simeq & e^{-2^{-m}n^{-1}(m)L},
\end{eqnarray}
for large $m$ and $L$.
Thus, we can write the time evolution (more precisely than in the
previous section) for the number of domains after $m$ typical
times, $n(m)$, which reads
\begin{eqnarray}
n(m+1) & = & \frac{1}{2} n(m)[1-q(m)] \\
       & = & \frac{1}{2} n(m)[1 - e^{-2^{-m}n^{-1}(m)L}].
\end{eqnarray}
If we define $\phi(m) = 2^{-m}n^{-1}L$ and if we consider $m$ as
a continuous variable, the last equation can be written as
\begin{equation}
\frac{d \phi}{dm} = \phi\, e^{-\phi},
\end{equation}
whose solution can be written in an implicit form
by means of the exponential integral function ${\rm Ei}(x)$
\begin{equation}
{\rm Ei}(\phi(m)) -{\rm Ei}(\phi(m_0))=m-m_0
\end{equation}
There is a problem of treating the initial conditions properly. It
is tempting to take the initial condition $m_0=1$ and therefore
have formally $\phi(m_0)=1/2$. However, it is not a prudent
approximation to assume the variable $m$ as continuous for very
small values. Taking the the initial condition for  $m_0 > 1$ we
do not know how to compute reliably the value $\phi(m_0)$.
However, we found that with $m_0=3$ and $n(3)=L/4$, i.e.
$\phi(3)=1/2$ we match well the numerical simulations data. Thus,
we have the equation ${\rm Ei}(\phi(m))= m - 3 +{\rm Ei}(1/2)$,
where ${\rm Ei}(1/2)=0.454...$ For the estimated rank $R(L)$, we
have $R(L) \simeq 1+2^{-m^{*}}L$, for the assumptions (i). Let us
define the reduced variable $\rho(L)=R(L)-1$, $\rho(L) \in [0,L-1]
$ so that we can write $\rho(L) = \phi(m^{*})$. For $\phi(m^{*})$
we have the equation ${\rm Ei}(\phi(m^{*}))= m^{*} - 3 +{\rm
Ei}(1/2)$, thus
%
%
%
\begin{equation}
{\rm Ei}(\rho(L))+\frac{\ln \rho(L)}{\ln 2}= \frac{\ln L}{\ln 2} - 3 +{\rm Ei}(1/2)
\label{eq:forphic}
\end{equation}
For $\rho(L)\to\infty$ we have $\ln({\rm Ei}(\rho(L))+\ln \rho(L)/\ln
2)\simeq \rho(L)$ (we checked by plotting the functions that it holds well
for $\rho(L)$ larger than $\simeq 30$), so we end up with the conclusion that
asymptotically for $L\to\infty$ the rank behaves like
\begin{equation}
\rho(L)\simeq \ln\ln L\quad .
\end{equation}
However, this asymptotic regime is reached only for extremely large
systems. Rank $\rho(L)$ larger than about 30 means $L$ larger than about
$10^{10^{12}}$: a number beyond any imaginable application. For smaller
sizes, up to about $L=1000$, we found approximately, by expanding the
LHS of equation (\ref{eq:forphic}) in Taylor series around $\rho(L)=1$ and
by replacing $\rho(L)$ by $R(L)-1$,
\begin{equation} \label{rank}
R(L) \simeq 0.347\,\ln L+0.933 \ldots
\end{equation}
We may compare this result with numerical results and find an
excellent agreement, as shows fig. \ref{fig2}.

\section*{Improving the filter}
In order to improve the performance of the filter, one could
follow the suggestions coming from the Shannon information theory
\cite{Sh}, where the addition of redundant information (e.g., by
repeating the transmission of the message) helps in recovering the
source information. In the same approach, we could build a number
$\alpha$ of replicas of the initial chain, in which each site
corresponds to a different value. Then, by linking the $\alpha$
chains together, keeping periodic boundary condition, we obtain a
chain of length $L'=\alpha L$. In the same Shannon's spirit, we
could wonder if there exists a finite value $\alpha_{c}$of
$\alpha$ such that the inefficiency of the filter decreases to its
minimal value, and the final selected value is (almost) always the
highest one. In the new $L'$-chain, the first $\alpha$ places in
the ranking of the values are occupied by the $\alpha$ replicas.
Therefore, if $R(L') < \alpha$, the selected value is the best one
attached to a replica of the original site. We have the following
condition for $\alpha_{c}$,
\begin{equation}
\frac{1}{3}\log{(\alpha_{c}L)} \leq \alpha_{c},
\end{equation}
which has a finite solution $\alpha(L)$, for all values of $L$,
which increases with $L$.
As a consequence, however, the number of matches that have to be
done to reach a steady state is increased as the number of
sites grows from $L$ to $L'=\alpha L$.
We denote by $M(l)$ the total number of matches needed to reach
a steady state in a chain of $l$ sites.
If we assume that each match costs a unit time, the quantity
$C=\frac{M(L)}/{M(\alpha_{c}(L)L)}$, is analogous to the inverse
of the information capacity in Shannon's theory, which
decrease with increasing redundancy.
In the original chain of length $L$, at each time step $L$ matches
are made. Then, from \ref{t*_L}, we have $M(l) \simeq l^{3}$.
>From the definition of $C$, we get the relation
\begin{equation}
C \simeq \alpha_{c}(L).
\end{equation}
\section*{Conclusions}

We have introduced a toy model of search engine, i.e. an algorithm
which elects a {\em best} element in a large set, according to
some relevance criterion. 
Such a device is gaining a growing importance in the current 
information-based economy: nowadays, gathering large 
amounts of informations is a widely affordable task; 
on the other hand, selecting, examining and 
judging information requires a huge (and increasing) processing 
capacity, though it is necessary to exploit such available 
information.

These algorithms face two main problems. 
First, the exact relevance of an item (an information, a 
webpage, a people) in the most common situations can be 
measured only within a certain degree 
of uncertainty. 
Only an infinite available time would allow a deep 
knowledge about the items.
Second, the results have to be presented in a user-friendly manner,
e.g. in ranking order. In the extreme case, the algorithm 
will even yield only the {\em best} selected item.

This puzzle recalls the one solved by pioneers of Information Theory,
who dealt with the challenge of recovering the original message
transmitted through a noisy channel, by knowing only the
corrupted received message.
They established that, even with a finite information capacity,
one is able to achieve error-free communication, though
they rarely constructed such algorithms. 
Analogously, nowadays one looks for methods that are able
to order large sets of items with respect to their intrinsic 
relevance, without having a full knowledge about them.

Accordingly, we assumed that the quality of each 
element can not be measured directly, and one can only compare 
elements pairwise. 
Each element is put on a site of a linear chain with
periodic boundary conditions, and can spread over neighboring
sites, thus creating domains. 
The ``search engine'' stops when a domain occupies the whole lattice
and the value attached to the domain is selected as the 
{\em best} one.
 
The model is approached by applying methods issued from a 
directed polymers field, since its properties can be investigated 
by focusing on the domain walls dynamics. 
We analytically computed the search time and the inefficiency 
of the ``filter'', and verified our results by numerical simulations.

Most interestingly, we found that the error made by the filter 
(the intrinsic ranking of the elected item) grows only 
logarithmically with respect to the number of items.
Moreover, we determined the minimal redundancy to be added
to the filter in order to achieve full efficiency, i.e. to
always select the intrinsically most relevant item in the 
set. 
In the analogy with classical Information Theory,
this would correspond to the well-known Shannon limit.

\section*{Acknowledgments}
The authors thank P. Laureti, P. De Los Rios, Y. Manida and Y. Pismak for
useful discussions.
This work was supported by the Grant Agency of the Czech Republic,
grant project No. 202/01/1091, and by the Swiss National Fund, Grant 
No. 20-61470.00, and by the Grant Agency of the Czech Republic, 
Grant No. 202/01/1091.
F. Slanina acknowledges the financial support from the University of
Fribourg, Switzerland.

\end{document}